\documentclass[preprint,11pt]{aastex}
\usepackage{ifthen}
\newboolean{apj}
\setboolean{apj}{false} 
\newboolean{color}
\setboolean{color}{false}  

\ifthenelse{\boolean{apj}}{\usepackage{emulateapj5}}{}

\newcommand{\Lya}{Ly$\alpha$}
\newcommand{\lya}{Ly$\alpha$}

\newcommand{\Dave}{Dav\'{e}}
\newcommand{\MK}{\,\mbox{K}}
\newcommand{\Merg}{\,\mbox{erg}}
\newcommand{\Mcm}{\,\mbox{cm}}
\newcommand{\Ms}{\,\mbox{s}}
\newcommand{\Myr}{\,\mbox{yr}}  
\newcommand{\MMpc}{\,\mbox{Mpc}}
\newcommand{\Mkpc}{\,\mbox{kpc}}
\newcommand{\Mkms}{\,\mbox{km}\,\mbox{s}^{-1}}
\newcommand{\msun}{M_{\sun}}
\newcommand{\tsim}{\sim\!}
\newcommand{\esfunit}{\Merg \Ms^{-1} \msun^{-1} \Myr}
\newcommand{\taulast}{\tau_{\it last}}
\newcommand{\cdu}{\Mcm^{-2}}

\begin{document}

\title{Cooling Radiation and the Lyman-alpha Luminosity of Forming Galaxies}

\author{Mark A. Fardal, Neal Katz}
\affil{Astronomy Department, University of Massachusetts, Amherst, MA 01003}
\author{Jeffrey P. Gardner}
\affil{Department of Astronomy, University of Washington, Seattle, WA 98195}
\author{Lars Hernquist}
\affil{Department of Astronomy, Harvard University, Cambridge, MA 02138}
\author{David H. Weinberg}
\affil{Astronomy Department, Ohio State University, Columbus, OH 43210}
\author{Romeel Dav\'e}
\affil{Astrophysical Sciences, Princeton University, Princeton, NJ 08544}

\begin{abstract}
We examine the cooling radiation from forming galaxies in hydrodynamic
simulations of the LCDM model (cold dark matter with a cosmological constant),
focusing on the Ly$\alpha$ line luminosities of high-redshift systems.
Primordial composition gas condenses within dark matter potential wells,
forming objects with masses and sizes comparable to the luminous regions of
observed galaxies.  As expected, the energy radiated in this process is 
comparable to the gravitational binding energy of the baryons, and the total 
cooling luminosity of the galaxy population peaks at $z \approx 2$.  However,
in contrast to the classical picture of gas cooling from the $\sim 10^6\;$K
virial temperature of a typical dark matter halo, we find that most of the
cooling radiation is emitted by gas with $T<20,000\;$K.  As a consequence,
roughly 50\% of this cooling radiation emerges in the Ly$\alpha$ line.  While
a galaxy's cooling luminosity is usually smaller than the ionizing continuum
luminosity of its young stars, the two are comparable in the most massive
systems, and the cooling radiation is produced at larger radii, where the
Ly$\alpha$ photons are less likely to be extinguished by dust.  We suggest, in
particular, that cooling radiation could explain the two large 
($\sim 100\;$kpc), luminous ($L_{{\rm Ly}\alpha} \sim 10^{44} \Merg \Ms^{-1}$)
``blobs'' of Ly$\alpha$ emission found in Steidel et al.'s (1999) narrow band
survey of a $z=3$ proto-cluster.  Our simulations predict objects of the 
observed luminosity at about the right space density, and radiative transfer
effects can account for the observed sizes and line widths.  We discuss 
observable tests of this hypothesis for the nature of the Ly$\alpha$ blobs,
and we present predictions for the contribution of cooling radiation to the
Ly$\alpha$ luminosity function of galaxies
as a function of redshift.

\end{abstract}

\keywords{cosmology: theory --- galaxies: formation --- galaxies: ISM}


\newcommand{\coolfracfig}{
\begin{figure*}
\ifthenelse{\boolean{color}}{\plotone{figs/cooltemp_L144.eps}}{\plotone{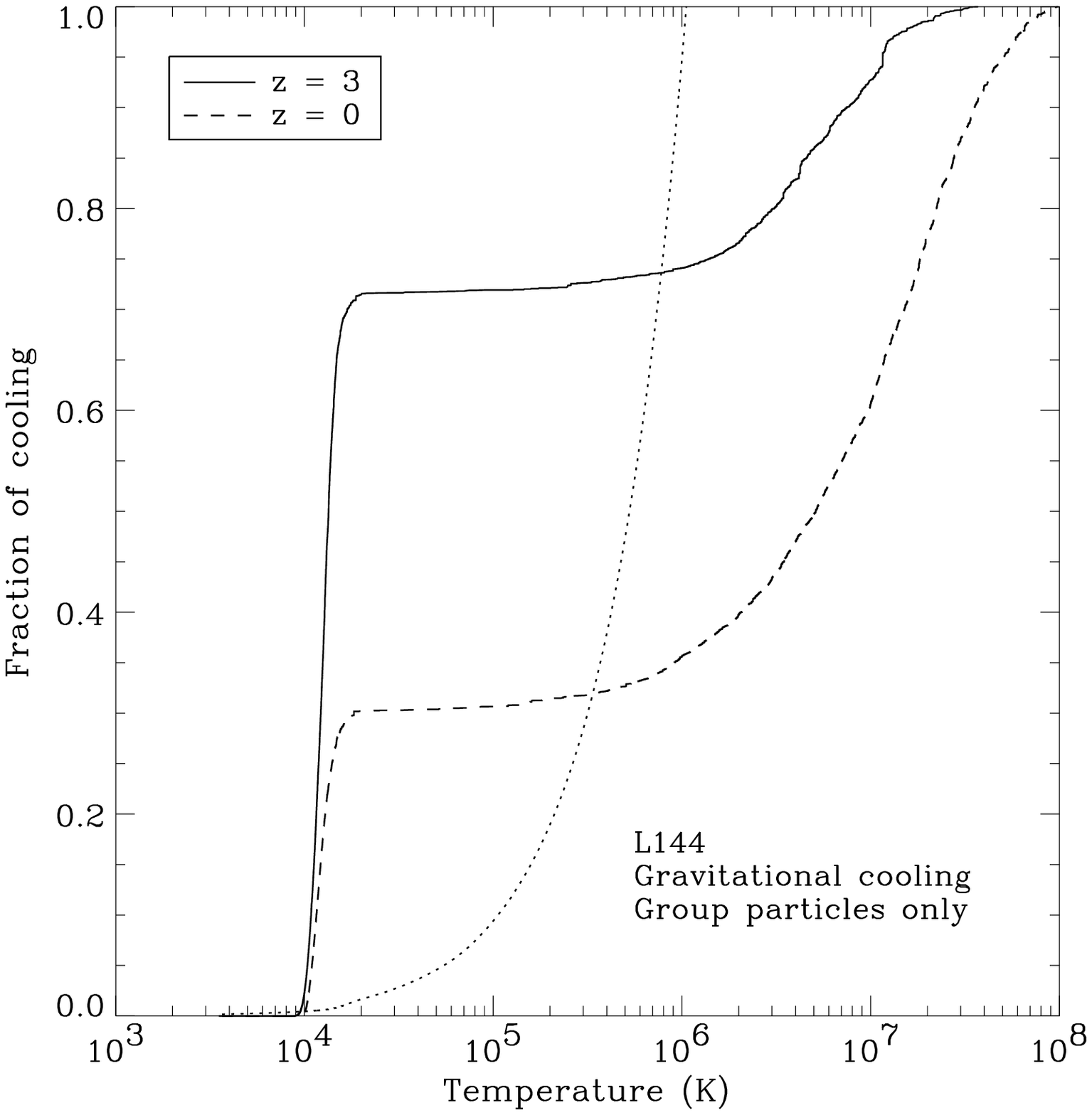}}
\caption{
\label{fig.coolfrac}
Cumulative distribution of the cooling radiation as a function of the
temperature of the emitting gas, for the L144 simulation at $z=3$
(solid) and $z=0$ (dashed).  The
supernova contributions to the cooling are omitted.  The 
dotted line shows the distribution expected for gas that cools from an
initial temperature of $10^6 \MK$.  Since the gas is fully ionized in
H and He down to $\tsim 8 \times 10^4 \MK$, this curve is
approximately $T/(10^6 \MK)$ over most of its range
of significance. }
\end{figure*}
}

\newcommand{\lfcumfig}{
\begin{figure*}
\ifthenelse{\boolean{color}}{\plotone{figs/lfcum_src3.eps}}{\plotone{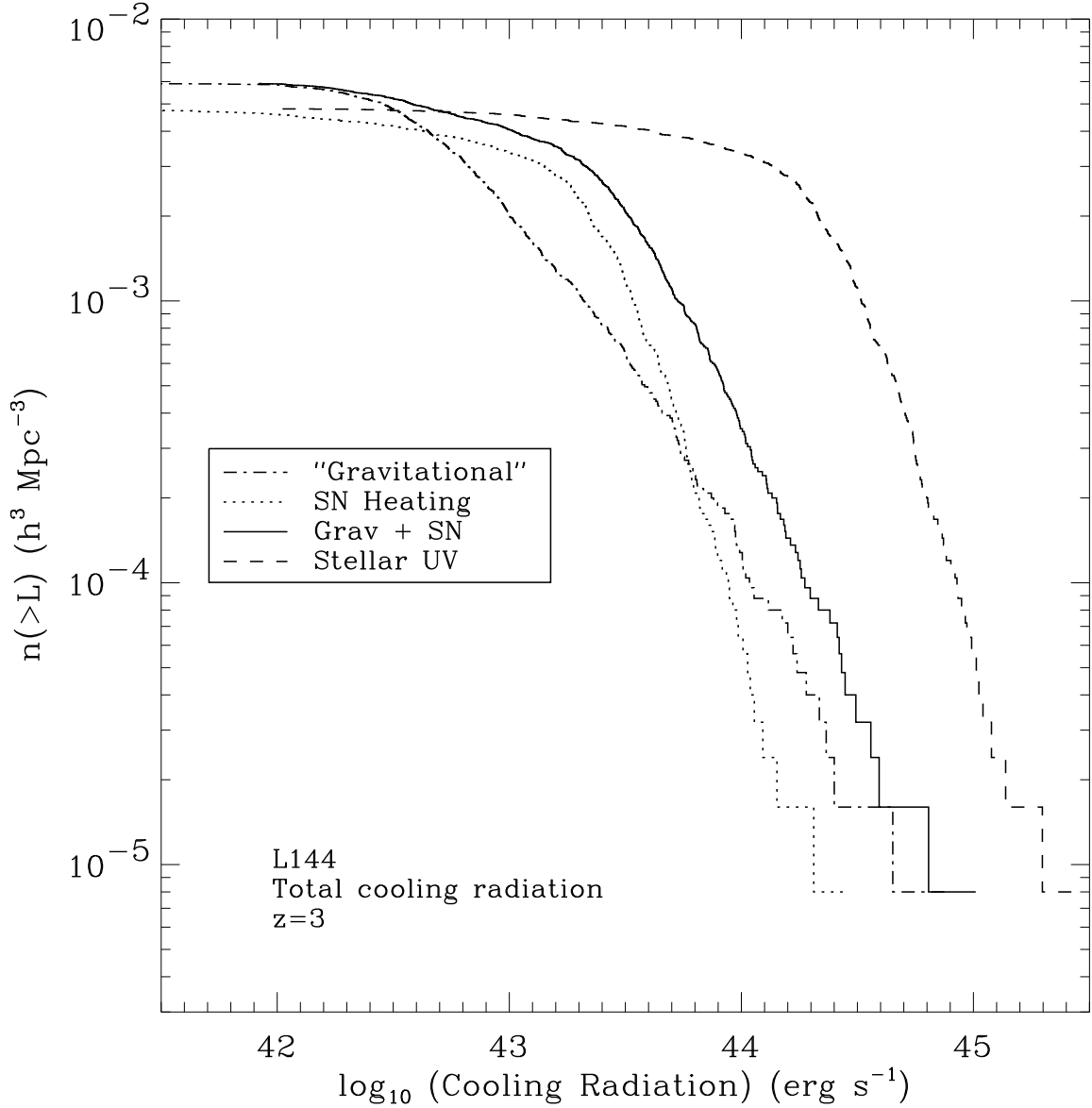}}
\caption{
\label{fig.lfcum_src}
Luminosity functions based on the different sources of energy input to
the gas.  The gravitational cooling luminosity function of groups in
the L144 simulation at $z = 3$ is plotted as the dot-dashed line.  The
contributions from supernova heating and stellar Lyman continuum are
plotted as the dotted and dashed lines respectively.  The sum of
gravitational and supernova inputs are plotted as the solid curve.
The turnover in the curves at a density of $3 \times 10^{-3} \, h^3
\MMpc^{-3}$ is an artifact of limited resolution.  All luminosity
functions in this paper are plotted using comoving densities.  }
\end{figure*}
}

\newcommand{\lyalfzevlnfig}{
\begin{figure*}
\ifthenelse{\boolean{color}}{\plotone{figs/lyalfcum_zevln.eps}}{\plotone{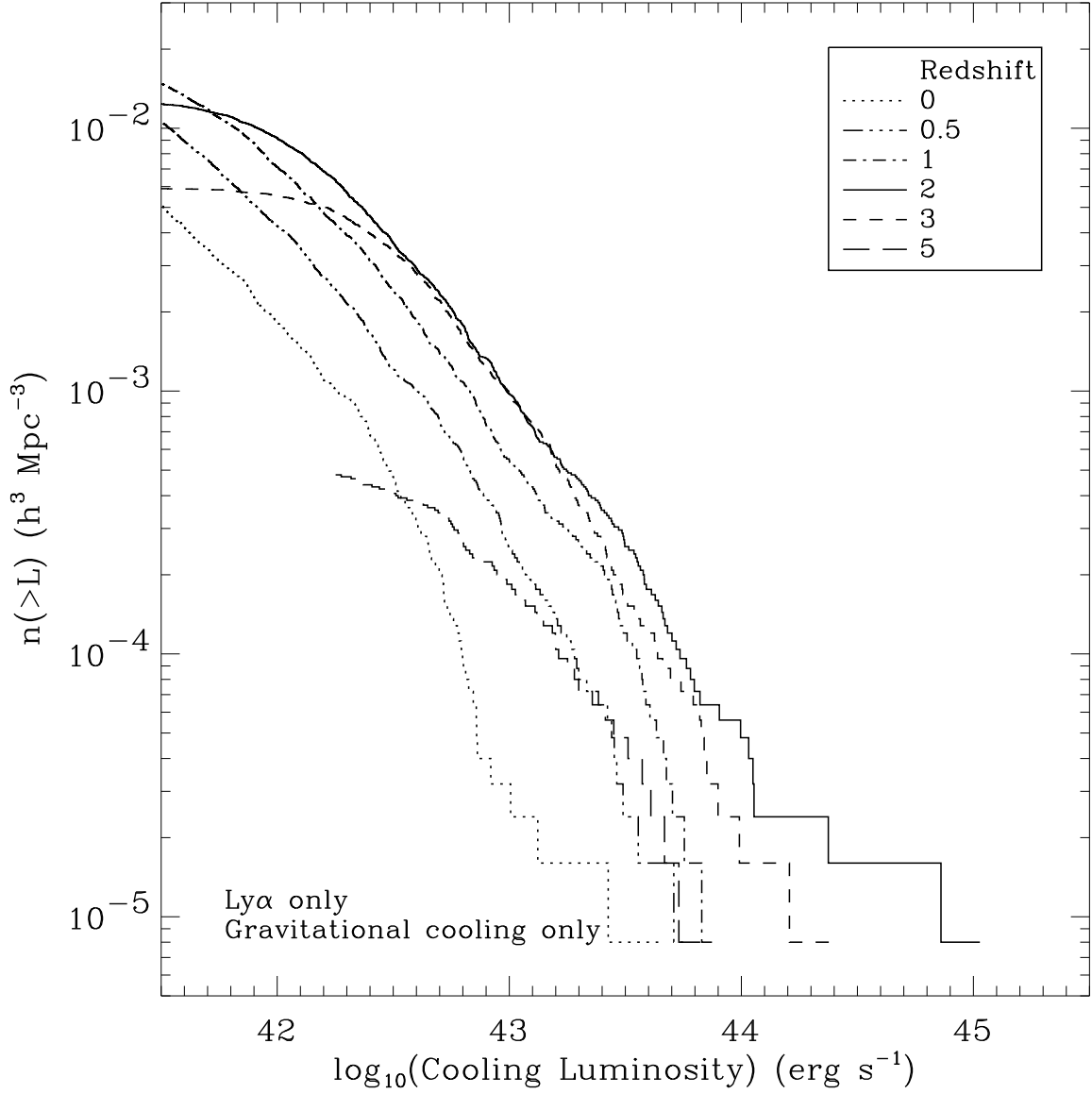}}
\caption{
\label{fig.lyalf_zevln}
Evolution of the cumulative \Lya\ cooling luminosity function at several
redshifts, for the L144 simulation. Only the contribution from gravitational sources
is included.}
\end{figure*}
}

\newcommand{\lyaimagefig}{
\begin{figure*}
\plotone{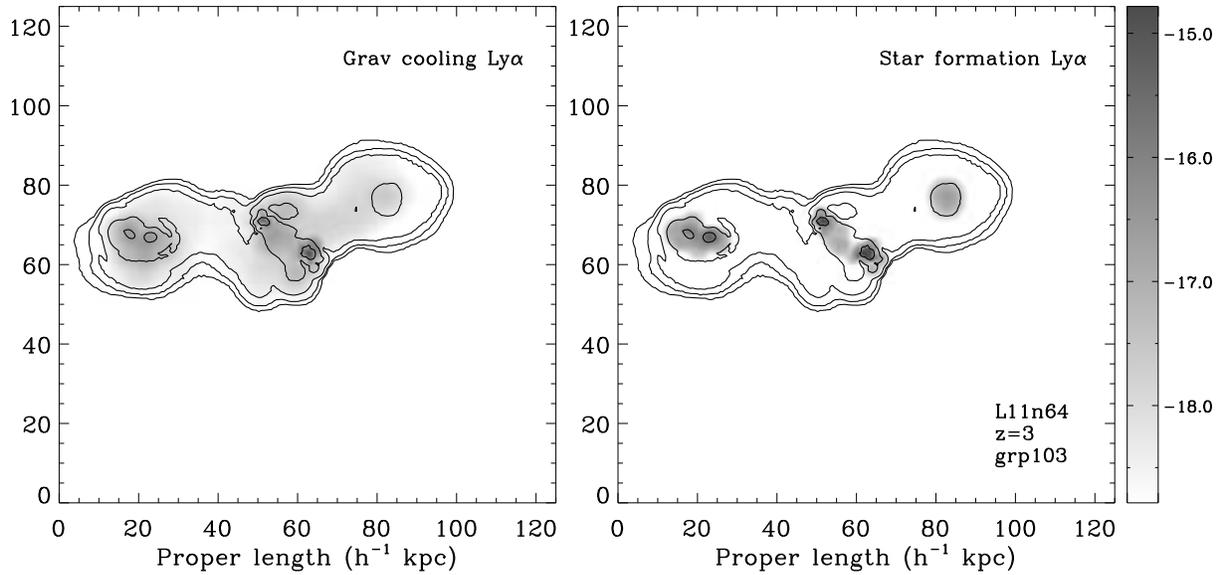}
\caption{
\label{fig.lyaimage}
An example of the \Lya\ radiation resulting from gravitational
cooling (left panel) and stellar photoionization  (right panel) 
in one of our groups at $z=3$.  
The example here is taken from our highest resolution simulation L11/64.
The emission is shown by the gray scale
images; the intensity scale is marked in terms of
$\log_{10}[I_{Ly\alpha}/(\Merg \Ms^{-1} \Mcm^{-2} \, \mbox{asec}^{-2})]$.
The neutral hydrogen column density is also shown
at contours of $\log_{10}(N_{\rm H I}/\Mcm^{-2}) = 18$, 19, 20, 21, and 22.
}
\end{figure*}
}

\newcommand{\extentfig}{
\begin{figure*}
\ifthenelse{\boolean{color}}{\plotone{figs/extent_L144_z3.eps}}{\plotone{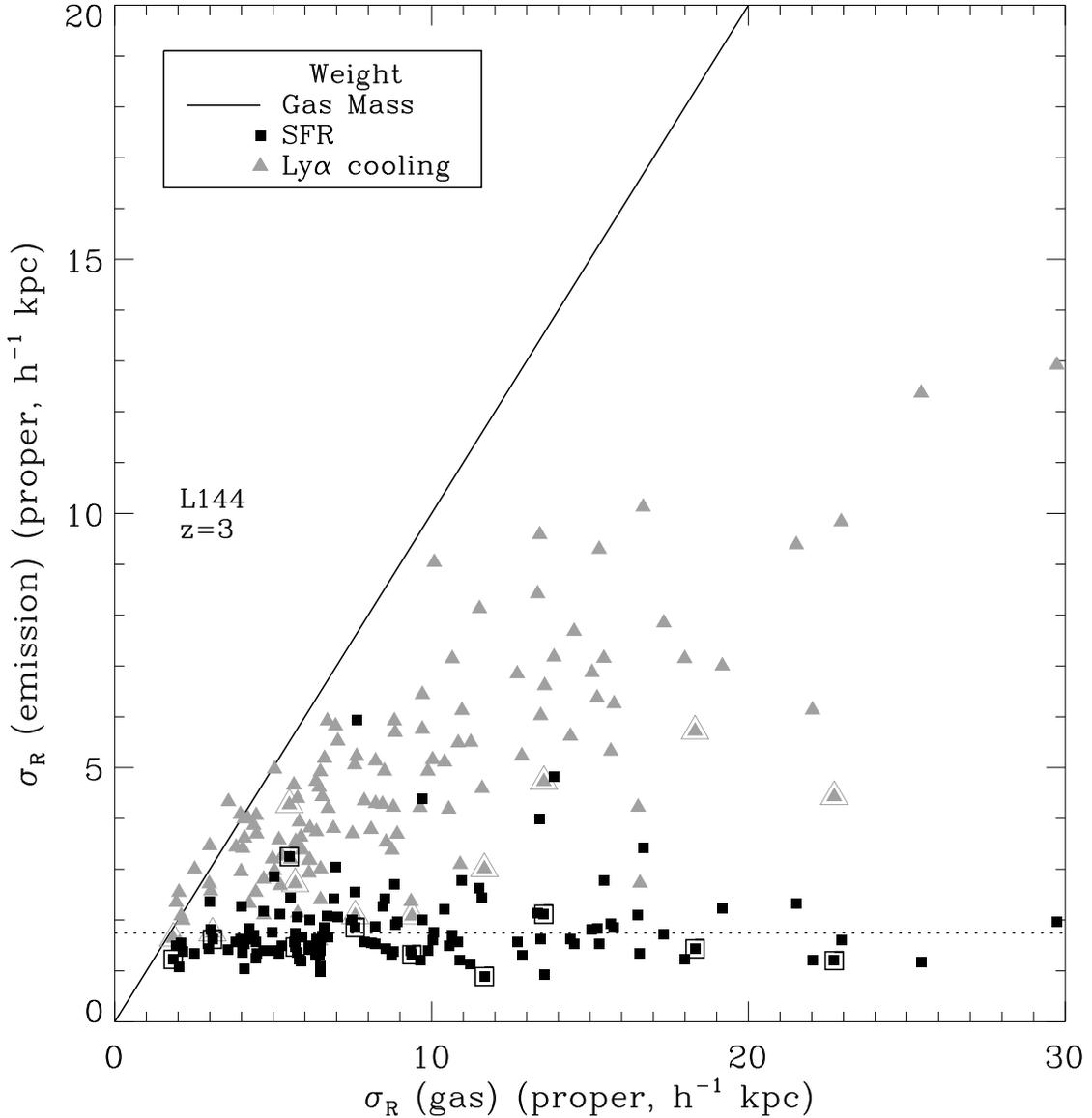}}
\caption{
\label{fig.extent}
Radial rms extent of the gas in the cooling groups at $z=3$, weighted
by several different factors.  The $x$-axis is weighted by the gas mass.
On the $y$-axis, the triangles are weighted by the Ly$\alpha$ emission
(considering only the ``gravitational'' cooling).  The squares are
weighted by the star formation rate, which is proportional to the
stellar ionizing emission.  The dotted line shows the gravitational
softening length.  The points would lie on the solid line if the emissivity
were proportional to the gas density.
Only groups with \Lya\ cooling of more than
$10^{43} \Merg \Ms^{-1}$ are shown; groups with \Lya\ cooling of more
than $5 \times 10^{43} \Merg \Ms^{-1}$ are shown surrounded by
outlines.}
\end{figure*}
}

\newcommand{\bmassfig}{
\begin{figure*}
\ifthenelse{\boolean{color}}{\plotone{figs/baryonmass_all_z3.eps}}{\plotone{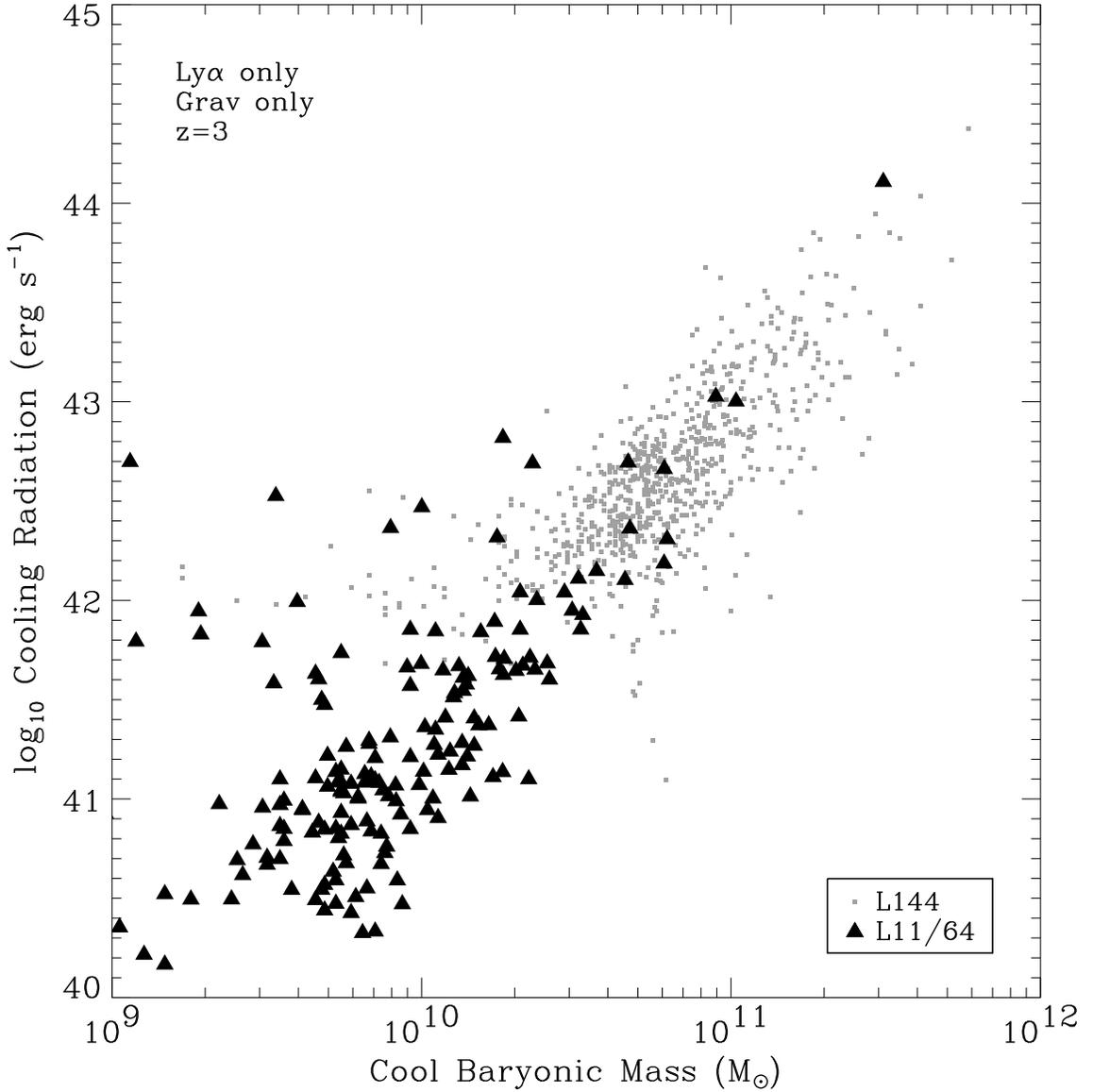}}
\caption{
\label{fig.bmass}
Cooling as a function of the ``galactic mass'', at $z=3$.
This is defined as the mass of the gas with temperature
$T < 3 \times 10^4 \MK$ and overdensity $\rho_g/\bar{\rho_g}>1000$,
plus the mass in
stars.  The cooling includes only the gravitational \Lya\ cooling.
The galaxies in the L144 simulation and the L11/64 simulation are
plotted as squares and triangles respectively.}
\end{figure*}
}

\newcommand{\sfrfig}{
\begin{figure*}
\ifthenelse{\boolean{color}}{\plotone{figs/sfr_z3.eps}}{\plotone{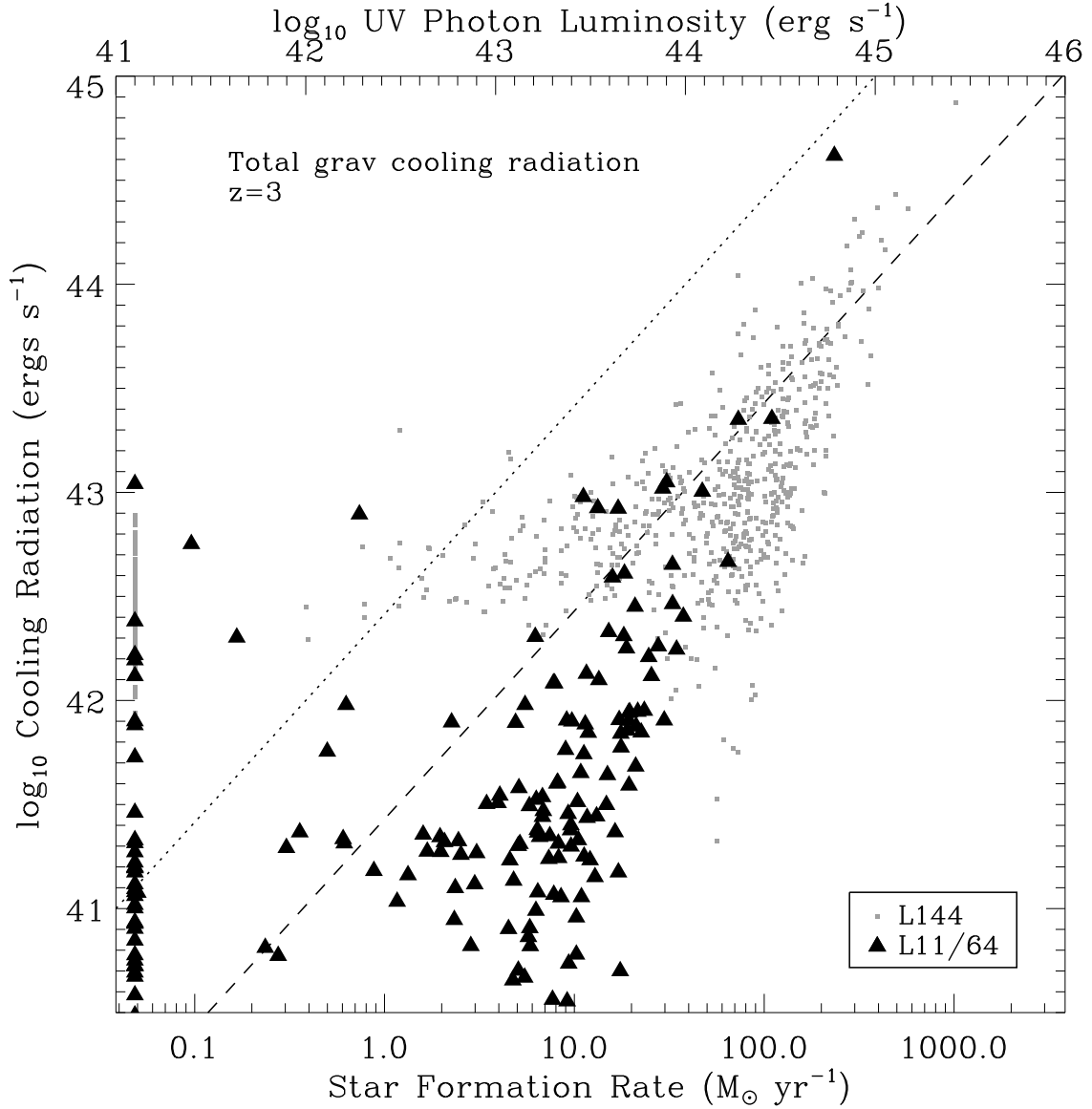}}
\caption{
\label{fig.sfr}
The total gravitational cooling radiation 
as a function of the star
formation rate, at $z=3$.  The galaxies in the L144 simulation and the
L11/64 simulation are plotted as squares and triangles respectively.
The Lyman continuum luminosity resulting from the star formation is
shown along the top axis.  The supernova heating resulting from star
formation is shown by the dashed line; galaxies lying on this line
have equal gravitational and supernova contributions to the
cooling radiation, and gravitational cooling dominates for points
above the line.  Galaxies on the dotted line would have gravitational
cooling radiation equal to the ionizing radiation output of their
young stars.  Galaxies with no star
formation are plotted at the extreme left. }
\end{figure*}
}

\newcommand{\epotfig}{
\begin{figure*}
\ifthenelse{\boolean{color}}{\plotone{figs/energy_z3.eps}}{\plotone{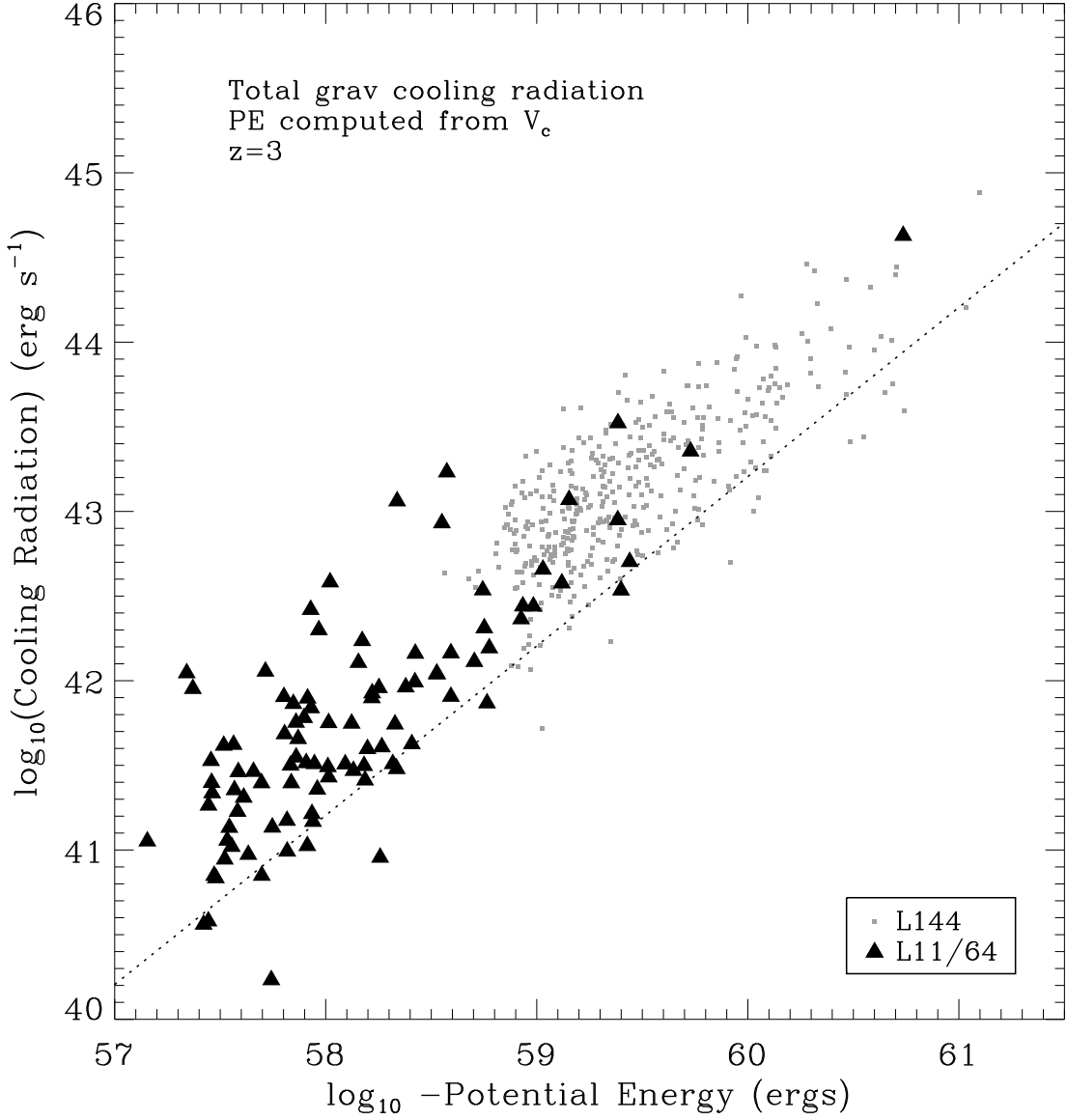}}
\caption{
\label{fig.epot}
Total gravitational cooling vs.
potential energy, at $z=3$.  The
galaxies in the L144 simulation and the L11/64 simulation are plotted
as squares and triangles respectively.  The potential energy here is
simply calculated from $U = M_b V_{circ}^2$, where $V_{circ}$ is the
measured circular velocity and $M_b$ is the baryonic mass.  The dotted
line shows the potential energy divided by the cosmic time. }
\end{figure*}
}

\newcommand{\compareresfig}{
\begin{figure*}
\plotone{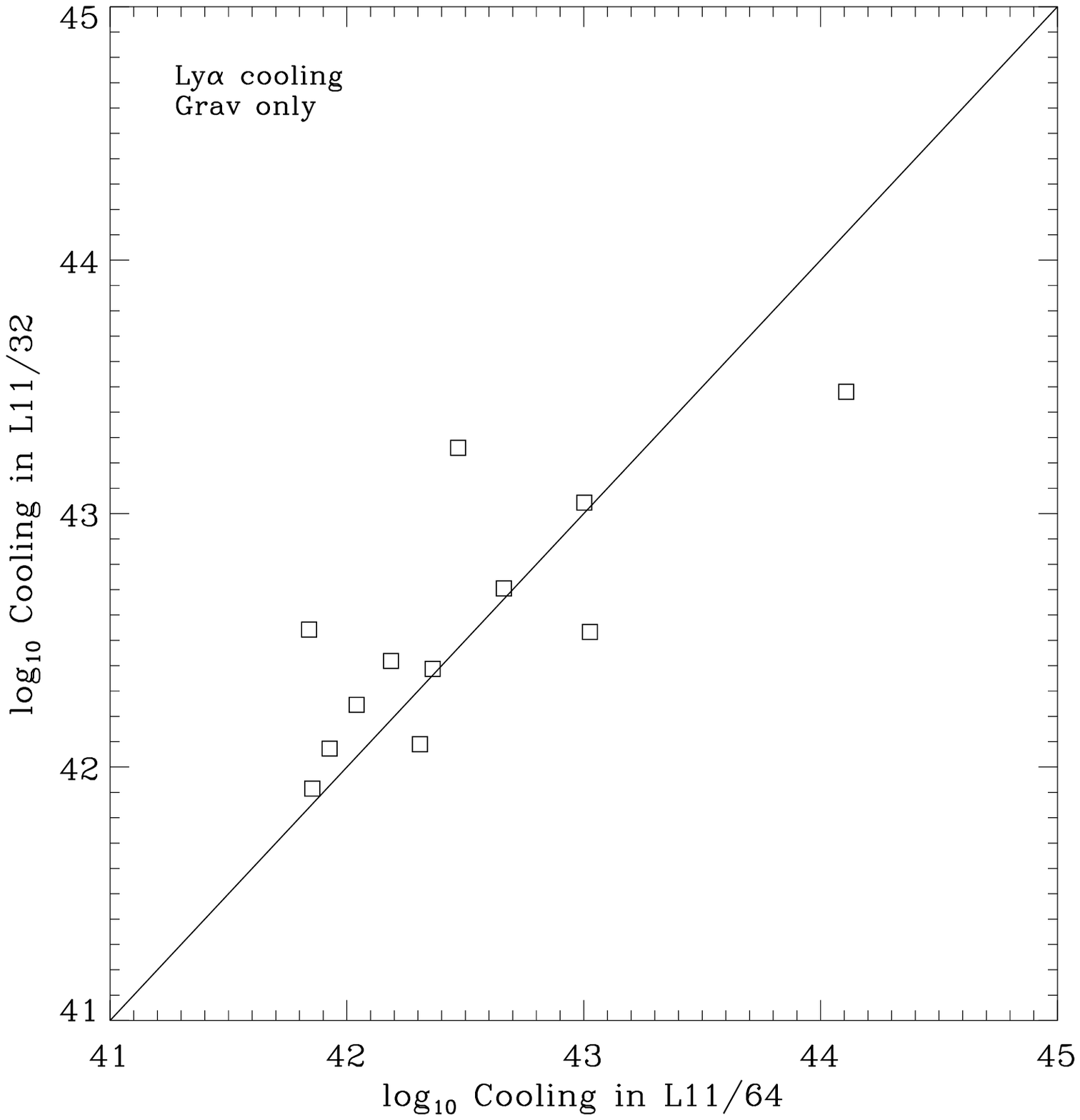}
\caption{
\label{fig.compare}
Cooling luminosity in \Lya\ at two resolutions at $z=3$. Galaxies from the
lower resolution L11/32 simulation, which has the same resolution as L144, are
plotted versus the same galaxies from the higher resolution,
 L11/64 simulation.
}
\end{figure*}
}

\section{INTRODUCTION}

Radiative cooling of gas in dark matter potential wells is an
essential element of the current theoretical understanding of galaxy formation.
As the baryons condense into tightly bound clumps at the centers of these
potential wells, they must radiate away the energy that they acquire through
compression and shock heating.  At first glance, it appears that this
radiation must be negligible compared to the radiation produced by stars:
the gravitational binding energy per unit mass is $\sim v^2 \sim 10^{-6}c^2$
even for a (high) internal velocity $v \sim 300\;\Mkms$, while the nuclear
energy per unit mass released in enriching gas to solar metallicity 
is $\sim 10^{-4}c^2$.  However, for primordial composition gas cooling
to $T \approx 10^4\;$K, all of the emitted radiation is either hydrogen
or helium line radiation (primarily hydrogen Ly$\alpha$) or continuum
radiation above the hydrogen ionization threshold.  Only the hottest stars
emit photons at these energies.  Furthermore, the cooling radiation is 
produced mainly at large radii, where it can escape the galaxy without
being absorbed by dust.  Finally, the kinetic energy of supernova explosions 
is deposited in dense, interstellar gas, where much of it can emerge in
the form of hydrogen line emission and ionizing continuum radiation.

Because of these factors, cooling radiation could make a significant
contribution to the X-ray, UV, and hydrogen line emission from young
galaxies.  In this paper, we examine this contribution using hydrodynamic
cosmological simulations.  Although many such simulations incorporate
radiative cooling, most studies have not followed the emitted radiation
in detail.  One notable exception is the work of Cen \& Ostriker
(\citeyear{cen92}, \citeyear{ostriker96}), who calculate the spectrum of the
background produced by cooling radiation, but these papers do not examine
the emission on a galaxy by galaxy basis.
Closer to the focus of this paper is the work of \citet{katz92},
which examined the formation of a single, massive galaxy, finding
that the amount of gravitational energy radiated in atomic lines 
was comparable to the amount of energy injected by supernovae.
Here we examine the cooling radiation from the whole population of
forming galaxies, though the individual objects in our simulations
are not as well resolved as that of \citet{katz92}.

The other main approach to theoretical modeling of galaxy formation
utilizes semi-analytic methods, in the tradition of \citet{white91}.
These methods rely on simplifying assumptions about gravitational
collapse and gas dynamics --- in particular, they usually assume that 
infalling gas shock heats to the virial temperature of the dark matter halo,
before cooling to join the central object.  Our results below suggest
that this assumption may break down in the messy assembly process
that characterizes hierarchical galaxy formation.

The observational study of star formation in high-redshift galaxies
has accelerated in recent years, with progress in wavelength ranges
from the UV/optical \citep{steidel96,williams96} to the IR
\citep{elbaz99} to the sub-mm \citep{barger99}.  The present paper
is motivated mainly by recent progress in \Lya\ emission line searches.
Historically, this approach to finding high-redshift galaxies has
proved difficult \citep{pritchet94,thompson95}, in part because the 
\Lya\ emission associated with star-forming regions may be damped
by the combined effects of resonant scattering and dust absorption.  
However, \Lya\ emission line searches have recently begun to bear fruit,
thanks to improvements in sensitivity and efficiency
\citep{hu98,steidel99,rhoads00}.

One of the most striking results of these searches is the discovery by
\citet[hereafter S99]{steidel99} of two large regions of diffuse \Lya\
emission, in their narrow band observations of a galaxy proto-cluster
at $z=3.1$.  These ``blobs'' do not resemble the typical galaxies 
detected in the proto-cluster.  They have characteristic angular
sizes of 15'' and fluxes of
$\approx 1.3 \times 10^{-15} \Merg \Mcm^{-2} \Ms^{-1}$.  
For the cosmological model that we adopt in this paper
($\Omega_m=0.4$, $\Omega_\Lambda=0.6$),
these properties imply proper linear sizes of
$l \approx 75 \, h^{-1} \Mkpc$ and luminosities of
$L_{Ly\alpha} \approx 4.4 \times 10^{43} \, h^{-2} \Merg \Ms^{-1}$.\footnote{
  With $\Omega_m=1$, $\Omega_\Lambda=0$,
  these numbers become
$l \approx 55 \, h^{-1} \Mkpc$ and 
$L_{Ly\alpha} \approx 2.4 \times 10^{43} \, h^{-2} \Merg \Ms^{-1}$.  }
Some other workers may have observed similar objects.
\citet{francis99} found diffuse \Lya\ blobs in a cluster at $z=2.4$.
\citet{keel99} found three diffuse regions of \Lya\ emission with sizes of
$\tsim 40 h^{-1} \Mkpc$ and luminosities of 
$\tsim 10^{43} \, h^{-2} \Merg \Ms^{-1}$ (in our adopted cosmology),
also at $z=2.4$.  
These are somewhat smaller and less luminous than the blobs found by S99,
and some are located around known AGN, so they may or may not be related
phenomena.

Several explanations for these \Lya\ blobs have been proposed.
They could be caused by photoionization from a hidden central AGN (S99).
High-redshift radio galaxies often show emission line regions
with sizes and luminosities comparable to the S99 blobs, and the
absence of radio lobes could simply indicate a radio-quiet AGN; the
anisotropic emission of the standard unified AGN model would then
provide an explanation for the absence of nuclear optical emission.
Another possibility, also discussed in S99, is that the gas in the blobs is
photoionized by ultraviolet (UV) continuum radiation from young stars.
The required star formation rate is 
$\tsim 30 \, f_{\it esc}^{-1} \, \msun \Myr^{-1}$, where $f_{\it esc}$ is the
fraction of hydrogen ionizing radiation that escapes from the galaxy.
Low-redshift observations imply upper limits of 
$f_{\it esc} \lesssim 10\%$ \citep{leitherer95,hurwitz97}, so unless
these presumably gas-rich galaxies have much larger $f_{\it esc}$, the
required star formation rates are substantial.  However, current limits 
are quite weak, allowing rates $\lesssim 1000 \, \msun \Myr^{-1}$ 
\citep{steidel99}.
\citet{taniguchi00} have suggested that
the \Lya\ emission emerges from a galactic superwind that is powered by 
high star formation rates and extends to scales that are 
even larger than those of winds observed at low redshift \citep{heckman93}.
While the luminosities and sizes of these blobs can be
explained by wind models, the mechanism for channeling the wind
energy into \Lya\ emission is somewhat unclear.

Drawing on the results of our more general investigation,
we propose an alternative explanation for the \Lya\ blobs:
they represent gas that is radiating away its gravitational potential
energy as it settles into massive galaxies.  
We describe our numerical simulation techniques, with particular
attention to our treatment of radiative cooling, in \S2.
In \S3 we present our results for the cooling radiation and \Lya\
emission from forming galaxies.  In \S4 we discuss whether the cooling
radiation model can account for the observed properties of the \Lya\ blobs.
Section 5 discusses some of the numerical uncertainties in our calculations,
compares our model to a similar model proposed independently
by Haiman, Spaans, \& Quataert~(2000) using semi-analytic methods,
summarizes our results and discusses their implications
and predictions.

\section{SIMULATIONS}

We perform our simulations using the parallel version of the
cosmological N-body/hydrodynamic code TreeSPH
\citep{hk89,kwh96,dave97}, a code that unites smoothed particle
hydrodynamics (SPH) \citep{lucy77,gingold77} with the hierarchical
tree method for computing gravitational forces \citep{barnes86,h87}.
Dark matter, stars, and gas are all represented by particles;
collisionless material is influenced only by gravity, while gas is
subject to gravitational forces, pressure gradients, and shocks.  The
gas can also cool both radiatively, assuming primordial abundances,
and through Compton cooling.

TreeSPH is fully adaptive in both space and time.  In SPH, gas
properties are computed by averaging or ``smoothing'' over a fixed
number of neighboring particles, 32 in the calculations here. Hence
the smoothing lengths in TreeSPH decrease in collapsing
regions, in proportion to the interparticle separation, and in
underdense regions the smoothing lengths are larger.  TreeSPH allows
particles to have individual time steps according to their physical
state, so that the pace of the overall computation is not driven by
the small fraction of particles requiring the smallest time steps.
There is a maximum allowed timestep, called the system timestep, and all
particles are integrated with this step or one a power of two smaller.
The timestep criteria are detailed further in \citet{kwh96}
and \citet{quinn00};
we set the tolerance parameter $\eta$ to 0.4.

Since this paper concerns cooling radiation, it is important to
understand how we evolve the thermal energy. Usually one has two
choices: to integrate the thermal energy equation explicitly or
implicitly.  Integrating explicitly would require a timestep about
three times smaller than the cooling time, which is much smaller than what would
be required to integrate the dynamical equations, making the
calculation prohibitively expensive computationally.  To integrate the
equations implicitly involves inverting an $N_{gas}\times N_{gas}$
matrix several times per timestep.  With $N_{gas} > 3\times 10^6$ in
our simulations, this would also be prohibitively expensive.  Instead
we take an intermediate approach, solving the thermal energy equation
semi-implicitly as described in Hernquist and Katz (1989).
Briefly, we integrate the changes in the thermal energy caused by
shocks and pressure forces explicitly, while we integrate those caused
by radiative cooling implicitly.
Since the radiative cooling depends only on a particle's own temperature
and density, independent of other particles, no matrix inversion is
required, and the time scale of the non-radiative processes is
comparable to the other time scales that govern the gas timestep.

The semi-implicit approach guarantees that we integrate the thermal
energy equation in a stable way, but it does not guarantee accuracy.
Accuracy is maintained by keeping the size of the timestep reasonably
small.  We accomplish this in two ways.  First, we integrate the
thermal energy equation for all gas particles, independent of their
dynamical timestep, using a timestep that is one half the size of the
smallest dynamical timestep of any particle.  Second, we damp the cooling rate
so that no gas particle loses more than a given fraction of its
thermal energy in one timestep (see Katz \& Gunn 1991); i.e., we
slightly change the physics to make the numerical integration more
robust.  This could have the effect of making some regions temporarily
hotter than they would have been if these numerical compromises were
not made.  In practice, it just makes some regions take two or three
timesteps to cool instead of one, since these regions have such short
cooling times.  
This could change the temperature at which the energy is radiated, but
it should not change the total radiated energy by a large amount.

The simulations calculate radiative cooling processes assuming a
primordial composition gas and ionization equilibrium. These processes
include collisional excitation, radiative and dielectronic
recombination, collisional ionization, and bremsstrahlung.  Molecular
processes and metal lines are omitted, which means that radiative
cooling alone cannot reduce the temperature below $\tsim 10^4 \MK$.
The gas cooling rates we use for these processes are listed in
\citet[][see their figure 1]{kwh96}.  
However, the rate at which the gas radiates energy
is somewhat different from the rate at which it loses thermal energy:
the emitted recombination radiation includes both the
actual gas cooling and the atomic ionization potential,
while collisional ionization removes thermal energy but yields
no radiation at all.  A 
possible deficiency of the simulations is the omission of He~I line
cooling, but we have verified that it is unimportant compared to the
other processes in the simulations presented here.  

For the purposes of this paper, we are particularly interested in the
\Lya\ emission.  We assume that a fraction 0.68 of recombinations to
H~I produce a \Lya\ photon, appropriate for optically thick gas at
$10^4 \MK$; accounting for the relative photon energies this channels
0.49 of the recombination energy through \Lya.  In practice
this is usually a small contribution compared to the collisional
excitation.  Excitation of H~I by collisions with electrons can result
in either a \Lya\ photon or in 2-photon decay from the metastable
2~$^2$S state, as well as photons from higher series in some cases.
The ratio between \Lya\ and 2-photon decay is weakly dependent on temperature.
By averaging over one simulation, we find that a fraction 0.59 of the
H~I collisional excitation is channeled through \Lya\, with almost all
of the remaining energy in the 2-photon continuum.  

TreeSPH can also include a metagalactic ionizing UV background field.
TreeSPH does not include radiative transfer, so even dense regions,
which in reality should be self-shielded, are exposed to the full
background field. This results in a large energy exchange and
unphysical gas radiation from these regions.  Hence, we have
restricted our analysis to simulations performed without such a
background, which should yield reliable results for our purposes.

In these simulations we include star formation and supernova feedback
using the algorithm described in \citet{kwh96}.  The algorithm forms
stars in dense cool regions at a rate essentially controlled by the
dynamical supply of gas.  For each star formation event, supernova
energy is added to the surrounding particles as thermal
energy on a time scale of $2\times 10^7$ years.

In the version of our code used for these simulations, the thermal
energy input from supernovae is added at the beginning of each system
timestep, not continuously.  The system timesteps are always larger
than the cooling time in these dense star forming groups, so the
excess thermal energy has radiated away by the end of the step when we
output the system state.  This actually turns out to be quite helpful,
since the remaining radiation is caused by the radiating away of
gravitational energy and thus can be measured in isolation.  We
can easily recover the supernova thermal input because it is
proportional to the star formation rate.  Using a Miller-Scalo initial
mass function (IMF) with cutoffs at 0.1 and $100 \, \msun$, the heat
input from supernovae is $2.7 \times 10^{41} \esfunit$ when smoothed
over the neighboring particles (the exact value would be 
$2.5 \times 10^{41} \esfunit$ if there were no smoothing).
The drawback to this method is that we cannot easily compute the
distribution of the supernova energy within galaxies or the processes
by which it is radiated away.  Since the supernova energy is deposited
in the dense ISM over a length scale determined by our resolution, its
space and temperature distributions are somewhat suspect in any case,
and we will not attempt to reconstruct them.

In addition to cooling radiation, we compute the photoionizing
radiation from hot stars from the simulation outputs.
Using the code STARBURST99
\citep{leitherer99}, we find a Lyman continuum intensity of $2.6 \times 10^{42}
\esfunit$ for a Miller-Scalo IMF, assuming a mean Lyman continuum
photon energy of 1.4~Ryd.  For gas at $10^4 \MK$, there are 0.68 \Lya\ 
photons emitted per photoionization \citep{charlot93}, giving an
associated \Lya\ 
emissivity of $9.4 \times 10^{41} \esfunit$.  
Because the stars that produce photoionizing radiation are short-lived,
we can take the stellar emissivity to be proportional to the 
instantaneous star formation rate, which we compute from the
gas distribution.  We do not include this UV emission or stellar winds
as a source of feedback in the simulations.  It is likely to have
even less dynamical impact than supernova feedback, since any energy captured
by the gas is deposited in the densest, star forming regions
with moderate temperatures, where it
can be quickly radiated away.

To summarize, the three sources of radiative energy included in these
simulations that could result in \Lya\ radiation are gravitational
cooling, supernova feedback, and photoionizing input from hot stars.
These sources have distinct physical origins, our code follows them
separately, and we will keep them separate in the discussions and
figures.  When we refer to the ``total cooling'', we mean the
gravitational cooling summed over all radiative processes, not that we
are including the supernova or photoionizing input.

We calculate the cooling radiation at discrete output times with the
analysis program
TIPSY,\footnote{http://www-hpcc.astro.washington.edu/TSEGA/tools/tipsy/tipsy.html}
using the same cooling algorithm as TreeSPH.  To identify discrete objects
in the simulations, we use the program 
SKID.\footnote{http://www-hpcc.astro.washington.edu/TSEGA/tools/skid.html}
Because SKID defines gravitationally bound groups of particles, we
will refer to these objects as ``groups,'' though each such group 
actually represents a single galaxy.
SKID slides particles meeting density and temperature criteria along 
density gradients until they reach a local maximum.  We use a density cutoff
corresponding to the edge of a virialized isothermal halo, or a gas
density of
$\rho_g = (\Omega_b/\Omega_m) (\rho_{vir}/3)$, where the mean density
of a spherical virialized halo $\rho_{vir}$ is 178 at early times and
is given in general by \citet{kitayama96}.  As in \citet{weinberg99},
we also restrict  ourselves to groups that contain more than 64
baryonic particles, roughly the mass resolution limit of
the simulations.  Once we identify the groups, we can easily add up the
cooling radiation for all particles in the group.  Cooling radiation
can also be emitted outside these groups, but it proves to be a minor
contribution except at very early times; most of this additional
emission comes from groups excluded by our mass resolution
criterion rather than from low density gas.

All three of the simulations we discuss in this paper assume a 
$\Lambda$-dominated cold dark matter 
cosmological model with $\Omega_m = 0.4$, $\Omega_\Lambda = 0.6$, $h \equiv
H_0 / (100 \Mkms \MMpc^{-1}) = 0.65$, and a primeval spectral index $n=0.93$.
With the tensor mode contribution, normalizing to COBE using CMBFAST 
\citep{seljak96,zaldarriaga98},
implies a normalization
$\sigma_8=0.8$, which provides a good match to cluster
abundances \citep{white93}.
We use the \citet[][equation D28]{hu96} formulation of the
transfer function.  We adopt a baryonic density $\Omega_b = 0.02 \, h^{-2}$
consistent with the deuterium abundance in high redshift Lyman limit systems
\citep{burles97,burles98}
and with the opacity of the \Lya\ forest \citep{rauch97}.
All of our simulations model
a triply periodic cubical volume.

The main and largest simulation we discuss is the L144 simulation
\citep{dave00}, with $144^3$ gas and dark matter particles, a box
length of $50 \, h^{-1}$ comoving Mpc on a side, and a gravitational
softening length $\epsilon_{grav} = 7 \, h^{-1}$ comoving kpc
(equivalent Plummer softening). The nominal gas mass resolution is
$5.4 \times 10^{10} M_\odot$, corresponding to 64 gas particles. To
investigate the effects of our finite resolution we perform two
additional simulations.  The L11/64 simulation has a higher spatial and mass
resolution, so it must be run in a smaller volume, $11.1 \, h^{-1}$
comoving Mpc on a side.  It has $64^3$ gas and dark matter particles,
$\epsilon_{grav} = 3.5 \, h^{-1}$ comoving kpc, and a gas mass
resolution of $6.8 \times 10^9 M_\odot$.  The L11/32 simulation is identical
to the L11/64 simulation, with the same initial phases, except that it uses
$32^3$ particles of each type and has the same resolution as the
L144 simulation.  In all the simulations the nominal spatial resolution in
physical units is $\sim 2\epsilon_{grav}/(1+z)$.

\section{COOLING RADIATION AND THE LY$\alpha$ LUMINOSITY FUNCTION}

Figure~\ref{fig.coolfrac} displays the temperature distribution of
the total gravitational cooling radiation (all radiative processes),
in the L144 simulation at $z=3$ and $z=0$.  In the 
conventional theoretical sketch of galaxy formation (e.g., 
\citealt{white78}; \citealt{white91}), gas that falls into dark
matter halos is shock heated to the virial temperature 
$T_{vir} = (\mu m_p / 2k) v_c^2 \approx 10^6\MK (v_c/165 \Mkms)^2$,
then cools and settles into the central galaxy.
The dotted line in Figure~\ref{fig.coolfrac} shows the temperature
distribution of cooling radiation expected for gas cooling from
an initial temperature of $10^6\MK$.  The numerical
simulation results paint a very different picture.
A large fraction of the cooling radiation, 75\% at $z=3$ and
30\% at $z=0$, comes from gas with $10^4\MK < T < 2\times 10^4\MK$.
Most of the remaining radiation comes from much hotter gas,
with $10^6\MK < T < 10^8\MK$.

\ifthenelse{\boolean{apj}}{\coolfracfig}{}

The lack of cooling radiation from gas with
$2\times 10^4\MK < T < 10^6\MK$ cannot be a result of gas
``moving quickly'' across this temperature range, since even
if it did, it would still have to radiate away its thermal energy
in order to reach $2 \times 10^4\MK$.  Instead, Figure~\ref{fig.coolfrac}
implies that most of the gas that cools into galaxies is never heated
to the virial temperature of a galaxy-mass dark halo.
This conclusion accords with that of \citet{kay00}, who find,
based on similar sorts of simulations, that only $\sim 10\%$ of
SPH particles that end up in galaxies were ever heated above $10^5\MK$.
We find qualitatively similar results in our own simulations,
but we have not examined particle temperature trajectories with the
high (single timestep) time resolution used by \citet{kay00}.

This physical result has major implications for the spectrum of cooling
radiation from forming galaxies, since much of the energy from neutral
gas at $T \sim 10^4\MK$ emerges in the \lya\ line as a result of
collisional excitation (see \citealt{kwh96}, figure 1).
The hot gas cooling, on the other hand, is dominated by bremsstrahlung.
At $z=3$, \lya\ emission accounts for 43\% of the cooling radiation
in the simulation, with 29\% emerging in HI 2-photon emission, 19\%
in bremsstrahlung, and only 8\% in other radiative processes.
If we assumed a metal-enriched intergalactic medium instead of
primordial composition, then more of the hot gas might be able
to cool, boosting the fraction of radiation from $10^5\MK < T < 10^7\MK$,
but the large emissivity of lower temperature gas would remain.

Our focus in this paper is
the cooling radiation, and especially the \lya\ emission, associated
with individual galaxies.  In Figure~\ref{fig.lfcum_src}, 
the dot-dashed line shows the cumulative luminosity function of
the total gravitational cooling 
at $z=3$.  By ``total'' we mean that this 
includes all of the cooling radiation processes,
not just \Lya\ radiation.  The reradiated supernova energy is plotted
as the dotted line.  The gravitational energy available to a group
increases more quickly with mass than the star formation rate.  The
dominant source of the total cooling radiation thus changes over from
reradiated supernova energy to gravitational energy as the mass and
luminosity increase.  
The two processes combined give the luminosity function shown
by the solid line.
Finally, the dashed line shows the stellar UV
emissivity, calculated from the star formation rate in the galaxy.
For all luminosities, there is more
energy in the UV radiation produced by the massive stars than in the
cooling radiation.  
However, the UV radiation from young stars is likely to be heavily
absorbed by dust, and reradiated in the infrared.
The gravitational cooling radiation emerges from lower density gas
and is more likely to escape.  The supernova energy may also stand
a much better chance of escaping, as supernovae can destroy the dense
clouds responsible for the heaviest absorption and deposit their
kinetic energy in a more diffuse medium.  

\ifthenelse{\boolean{apj}}{\lfcumfig}{}

We plot the cumulative luminosity function of the \lya\ cooling radiation
alone in Figure~\ref{fig.lyalf_zevln}, at several redshifts.
Since the fraction of supernova energy converted into \Lya\ 
is uncertain, we include only
gravitational sources of cooling in this plot.
Figure~\ref{fig.lfcum_src} shows that gravitational cooling
dominates in the most luminous objects.
The number of highly luminous objects reaches a peak at $z=2$,
and declines thereafter.
At $z=3$ there are $\tsim 4 \times 10^{-5} \, h^3 \MMpc^{-3}$ objects 
in the simulation with \Lya\ luminosities greater than $3 \times
10^{43} \, h^{-2} \Merg \Ms^{-1}$, comparable to the `blobs'' of S99.

\ifthenelse{\boolean{apj}}{\lyalfzevlnfig}{}

In the left hand panel of Figure~\ref{fig.lyaimage}, we show a map of
the \Lya\ emission contributed by gravitational energy in one of our
simulated groups, and in the right hand panel we plot the stellar
photoionization using a conversion factor of star formation to
recombination-induced \Lya\ of $6.6\times 10^{41} \esfunit$.  The
emission is shown by the gray scale images, with the intensity scale 
marked in terms of
$\log_{10}[I_{Ly\alpha}/(\Merg \Ms^{-1} \Mcm^{-2} \, \mbox{asec}^{-2})]$.
The emission from gravitational cooling 
is more spatially extended than the stellar emission.
Emission of cooling radiation from reradiated
supernova feedback would look like a scaled version of the right hand
panel, slightly smoothed by the feedback algorithm.  
We also plot contours of the neutral
hydrogen column density, calculated assuming an ionizing background of
$3 \times 10^{22} \Merg \Ms^{-1} \Mcm^{-2} \, \mbox{Hz}^{-1} \, \mbox{sr}^{-1}$
and correcting for self-shielding \citep{kwhm96}.  
Self-shielding accounts for the
rather sharp edges seen in the neutral gas (cf.\ Maloney 1993).
These emission plots ignore radiative transfer effects and dust extinction, 
both of which could greatly alter the observed appearance of these systems.
We discuss these issues in \S 4.

\ifthenelse{\boolean{apj}}{\lyaimagefig}{}

The spatial extent of the emission for the groups in the L144 simulation is
shown in Figure~\ref{fig.extent}.  The measure we use here is the root
mean square distance from the center of the group, weighted by the
Ly$\alpha$ emissivity.  The dotted line shows the gravitational
softening length; recall that the nominal spatial resolution is about
twice this length.  We plot the \lya\ extent against the rms extent of the gas
mass defined in a similar manner.  We also show the size 
weighted by star formation rate, which would represent
the size of the stellar UV emissivity region.
In all cases, the star formation in our simulations occurs in a small,
partially resolved region at the center of the group.  The \Lya\ emission
usually emerges over a region comparable to the size of the gas as a
whole and is well resolved numerically.  However, the most luminous
groups in the simulation tend to have more concentrated emission;
although there is weak emission at large radii, the typical rms sizes
are less than 10 kpc.  This size is much smaller than that of the
blobs observed by S99, a point we will return to in \S4.

\ifthenelse{\boolean{apj}}{\extentfig}{}

We plot the gravitational cooling in \Lya\ as a function of galactic
mass in Figure~\ref{fig.bmass}.  We define the galactic mass to be the
stellar mass plus the mass of the gas that is at least 1000 times
overdense and has temperature $T < 3\times 10^4 \MK$.  
To increase our dynamic range, we show both
the L144 simulation and the higher
resolution (but smaller volume) L11/64 simulation.  There is
a strong correlation between the emitted \Lya\ cooling radiation and
the galactic mass. In the L144 simulation, the galaxies with
high \lya\ cooling luminosities are all high mass
objects, many corresponding to $L_{\ast}$ galaxies or above.

\ifthenelse{\boolean{apj}}{\bmassfig}{}

In Figure~\ref{fig.sfr} we plot the total gravitational cooling radiation as a
function of the star formation rate at $z=3$.  Supernova feedback
would add an amount of cooling radiation shown by the dashed line.
The Lyman continuum luminosity associated with the stellar UV emission
is marked along the top axis.  
Star formation rates are underestimated in marginally resolved
systems, a numerical artifact that causes the spread in these 
scatter plots towards low star formation rates at relatively
low cooling luminosity.  Figure~\ref{fig.sfr} demonstrates on an
object-by-object basis the features seen in the luminosity
functions of Figure~\ref{fig.lfcum_src}.
Supernova cooling dominates gravitational cooling in low mass
objects, but gravitational cooling takes over at high masses.
The objects with the highest cooling luminosity also have
high star formation rates ($\mbox{SFR} \gtrsim 100 \, \msun \Myr^{-1}$).  
The stellar UV luminosity always exceeds the gravitational cooling
radiation (all points representing well resolved objects lie below the 
dotted line), but they are of similar magnitude in the most luminous
objects, so the cooling radiation would dominate if the stellar UV
is heavily extinguished by dust.

\ifthenelse{\boolean{apj}}{\sfrfig}{}

To examine the energetics of the cooling radiation, we
would like to define the gravitational potential energy available to
the baryons in the groups, but this definition is quite ambiguous.
Rather than being
symmetric, isolated entities, the groups are embedded in a complex,
filamentary, and clumpy structure.  We are interested in the energy
available to the baryons, but they interact with the dark matter in a
complex manner and energy is transferred between them.
Nevertheless, it is interesting to see whether the potential energy,
with the matter outside the group ignored, is a good predictor of
the cooling radiation.

In all of the previous figures, the groups defined by SKID consist of
baryonic particles only.  To study the gravitational energy, we must
redefine the groups to include dark matter particles as well, using a
density cutoff of $\rho_{tot} = \rho_{vir}/3$.  If we defined $U_{tot}$
to be the total potential energy of the group, and if the baryons and
dark matter were distributed identically with a baryon fraction
$f_b$, and if we assigned half of the baryon-dark matter interaction
energy to the baryon potential energy $U_b$, then we would have $U_b = f_b
U_{tot}$.  Another possibility would be to assign the entire interaction
energy to the baryons.  For example, if the cooling and collapse of
baryons to the center of the group takes place when the dark matter
halo is already assembled, then this latter definition is closer to
the amount of energy that the gas radiates.  As a practical matter,
$M_b V_c^2$, where $V_c$ is the circular velocity and $M_b$ is the
baryonic mass, usually lies between these two definitions.  Hence, we
take $M_b V_c^2$ as our estimate of the potential energy of the baryons.
One might object that, by the virial theorem, the energy available is
only $|U_b|/2$.  However, the virial theorem does {\em not} apply to
the baryons alone, as they are confined by the potential of the dark
matter.\footnote{In addition, the group as a whole is not an isolated
  system, and in our simulations the gravitational force is softened
  and thus not a power law; both of these facts also violate the
  conditions for the virial theorem to hold. }  
As a rule the sum of the kinetic and thermal energies 
falls short of even the smaller
definition of $|U_b|$ by a factor of 2 to 4.

A simple estimate of the cooling radiation is then $M_b V_c^2/t$, 
the potential energy divided
by the cosmic time.  At $z=3$ this prescription 
somewhat underestimates the cooling
from each group, as shown in Figure~\ref{fig.epot}, though the
discrepancy is not large
considering the ambiguity in our definition of $|U_b|$.
At $z=0$, this simple formula overestimates the cooling, indicating
that the cooling occurs preferentially at early times.
There could
be several reasons for this trend.  The cooling could be taking place less
effectively since the virial temperatures are rising; the rate of
accretion of matter into the group could have slowed; or the baryons
could become more effective at transferring their energy to the dark
halos.  We are unable to determine the answer from the discrete
outputs of the current simulations.  Tracking the evolution of
distinct groups as a function of time, as was done by \citet{katz92}
for a single galaxy, would give more insight into the energetics, but
we leave this investigation for future work.

\ifthenelse{\boolean{apj}}{\epotfig}{}

\section{MODELING THE LY$\alpha$ BLOBS}

So far, we have established that galaxy-like groups in our simulations
generically show large amounts of cooling in \Lya. Can this cooling
explain the ``blobs'' observed by S99?

The \Lya\ luminosities of some of our groups are as large as
those of the S99 blobs.
The number density of groups in our simulations with a \Lya\ luminosity
of $3 \times 10^{43} h^{-2} \Merg \Ms^{-1}$ is about 
$4 \times 10^{-5} \, h^3 \MMpc^{-3}$.  
At present we can make only crude estimates of the number density of
the observed \Lya\ blobs.  
 The volume examined by S99 has a size of
$8'.7 \times 8'.9$ and a depth of $\Delta \! z = 0.066$, or a comoving
volume of
$4200 h^{-3} \MMpc^3$ in our adopted cosmology.  As long as the blobs
are less abundant in other regions of space, they have a comoving
density of $< 5 \times 10^{-4} h^3 \MMpc^{-3}$.  A better estimate can
be obtained from S99's observation that this region is overabundant in
both Lyman break galaxies and in \Lya-emitting galaxies by a factor of
$\tsim 6$.  If the blobs are biased in the same way, they have a
density of $\tsim 8 \times 10^{-5} h^3 \MMpc^{-3}$, twice the density
we find.  If these blobs are associated with very massive objects,
they are probably more highly biased than Lyman break galaxies; hence
their actual number density is probably lower than this estimate.
With only two objects, which are probably correlated, the statistical
uncertainties are large.  So while the current constraints
are quite weak, the number density is consistent with that found in
our simulations.

The most luminous objects in our simulations typically have high mass
and are strongly clustered.  For example, the 100 most massive objects
in the L144 simulation have a comoving correlation length of 5.7
Mpc.  This strong clustering is consistent with the discovery
of the blobs in a proto-cluster, though of course the abundance of
blobs in blind fields is not well known.

As shown in Figures~\ref{fig.lyaimage} and \ref{fig.extent}, the length
scale over which most of the \Lya\ cooling is emitted in our
simulations is usually much smaller than the observed sizes of the
blobs in S99, though it is also larger than the typical size of the
star forming regions.  Hence, we must appeal to resonant
scattering to transport the \Lya\ photons to large radii.

The radiative transfer of \Lya\ in a static slab is a classic problem
\citep{adams72,adams75,neufeld90}.  The behavior of \Lya\ photons
depends upon the optical depth of the slab.  For intermediate optical
depths, the \Lya\ photons are spatially trapped, and they escape by
scattering into the tails of the Doppler distribution, where the slab
is optically thin.  For very large optical depths, the photons escape
by scattering into the damping wings, where they then perform a random
walk in both space and frequency.  The transition between these two
regimes occurs roughly where the damping wings become optically thick.
For gas with a temperature $T_4 \equiv T / (10^4 \MK)$, the total central
optical depth is
$\tau_0 = (N_{\it H I} / 1.7 \times 10^{13} \Mcm^{-2}) T_4^{-1/2}$,
and the damping constant is $a = 4.7 \times 10^{-4} T_4^{-1/2}$.  The
damping wings, which have the line profile
$\phi(x) = a / (\pi x^2)$,
become optically thick at 
$\tau_0 \approx 4 \times 10^4$, 
or 
$N_{\it H I} \approx 7 \times 10^{17} T_4^{1/2} \Mcm^{-2}$.
\footnote{Note that this column density is much less than what is
  conventionally referred to as a ``damped'' line, $N_{\it H I} \sim
  10^{20} \cdu$.}

Despite the extensive work on the slab problem, we have not found
calculations of the typical line-center optical depth at last
scattering ($\taulast$) in the literature. We can estimate $\taulast$
in two regimes.  In the case of intermediate optical depth, the
spatial diffusion is negligible and $\taulast \approx \tau_0/2$, the
optical depth to the slab center.  The photons escape with a
double-peaked profile with a typical frequency shift
$x \equiv \Delta \! \nu /  \Delta \! \nu_{\it Dop} = \sqrt{\ln \tau_0}$.
For large optical depths, the random excursion that
leads to escape takes place at a frequency 
$x_\ast \sim (a \tau_0 / 2 \sqrt{\pi})^{1/3}$
\citep{adams72}, for which the optical depth through the slab is 
$\tau_\ast = \tau_0 \phi(x_\ast) \sim (a \tau_0 / 2 \sqrt{\pi})^{1/3}$.
The last scattering of the photons in that excursion
occurs at optical depth $\tsim 1$
at that frequency, or a line-center optical depth of
$\taulast \sim \phi(x_\ast)^{-1} 
  \sim (\tau_0/2) (a \tau_0 / 2 \sqrt{\pi})^{-1/3}$.
Without performing Monte Carlo calculations, it is difficult to say
exactly where the transition between these regimes occurs; we will
take it to be $\tau_0 \approx 10^5$.
For very large columns, the \Lya\ photons are extinguished by dust.
This occurs for 
$N_{\it H I} \gtrsim 4 \times 10^{20} \cdu T_4^{1/2} \xi_{\it dust}^{-3/4}$ 
where $\xi_{\it dust}$ is the dust-to-gas ratio relative to
the Galactic value \citep{neufeld90}.
Even though galaxies at high redshift might have low metallicity,
star-forming regions should still be fairly rich in dust.  We 
somewhat arbitrarily take 
$\xi_{\it dust} = 0.1$ as a typical value.

Most of the neutral gas in our baryonic groups is at about $10^4 \MK$.
Hence, we might expect that \Lya\ photons emitted at 
$N \gtrsim 2 \times 10^{21} \cdu$
are absorbed by dust.  Photons emitted at 
$2 \times 10^{18} \cdu \lesssim N_{\it H I} \lesssim 2 \times 10^{21} \cdu$
are scattered in the line wings and finally escape in the range
$7 \times 10^{17} \cdu \lesssim N_{\it H I} \lesssim 7 \times 10^{19} \cdu$.
Finally, photons emitted at 
$N_{\it H I} \lesssim 2 \times 10^{18} \cdu$
are scattered mostly in place and eventually escape not far from their
region of formation.  

Although we are using one-dimensional models for these estimates, they
should be at least crudely applicable to our three-dimensional groups.
Applying this picture to Figure~\ref{fig.lyaimage}, we see that \Lya\ 
photons originating from stellar photoionization are likely to be
absorbed.  This is consistent with the strong absorption seen in
star-forming galaxies at high redshifts \citep{smail97}.
However, many of the photons from gravitational cooling may be
scattered outwards to the $10^{18} \Mcm^{-2}$ neutral hydrogen contour or
about $\tsim 30 h^{-1} \Mkpc$, with velocity widths up to
$300 \Mkms$.  In general we find that much of the gravitational cooling
is emitted outside the region of intense star formation.  Usually
$\tsim 20\%$ is generated outside our nominal dust cutoff, but
this fraction can be smaller or larger depending on the value we choose for
the cutoff.  A crucial ingredient in this argument is the presence of
reservoirs of neutral gas at large radii around the largest groups in
our simulations.  The velocities obtained are not quite as high as in
Figure~8 of S99, but the greatest velocity spread there comes 
from isolated knots and the velocity of the truly diffuse
emission is mostly unconstrained.

Clumping of the gas could make the escape of the photons easier and shrink
the apparent size of the \Lya-emitting regions \citep{neufeld91}.  The
presence of large bulk motions can greatly affect the line transfer,
probably explaining the moderate fraction of cases where photons escape
from star-forming regions \citep{kunth98,ahn98}.  However,
calculations using the velocity fields of our groups,
which are not well resolved in any case, are beyond the
scope of this paper.  We can conclude at this point only that our
model may be consistent with the observed sizes and velocity widths
of the \Lya\ blobs.

The cooling gas model for the blobs makes several testable
predictions.  Since the \Lya\ emission is expected to be scattered to
larger radii, emission in other lines should be more centrally
concentrated.  The neutral hydrogen column should be $\tsim 10^{18}
\cdu$ out to the radius of the \Lya\ blobs.  Since the \Lya\ is caused
mostly by collisional excitation rather than photoionization, the
H$\alpha$ flux should be quite small relative to \Lya\ ($\lesssim
2$\%).  These predictions are at odds with the results of
\citet{francis96} and \citet{francis97}, who report a detection of
H$\alpha$ and CIV in their three blobs and He~II Balmer-$\alpha$ in
two. In their brightest blob, labeled B1, H$\alpha$ has a similar
strength and distribution to \Lya.  This suggests that at least these
blobs are not due to gravitational cooling; this is concordant with
the apparent double-lobed morphology of B1 and red stellar colors of
the central galaxies, which suggest a AGN origin for these blobs.
H$\alpha$ unfortunately falls in the K-band only in the range $2.0 < z
< 2.6$, so it is important to search for other possible lines in the S99
blobs.

Another implication of our model is that large luminosities in diffuse \Lya\ 
emission are associated with massive objects with large star formation rates.
In fact, if Figure~\ref{fig.sfr} is taken at face value, the star
formation rates implied for the S99 blobs are $\tsim 100 \, \msun
\Myr^{-1}$.  However, the tight relation between cooling luminosity
and star formation rate is due to a similarly tight relation between
baryonic mass and star formation rate.  These
quantities may not be as correlated in real galaxies.  The more
fundamental prediction is that there is a massive galaxy at the heart of each
\Lya\ blob.

As mentioned before, to power the \Lya\ nebulosity with
Lyman continuum radiation from young stars, one needs
$\tsim 300 \, (f_{\it esc}/0.1)^{-1} \, \msun \Myr^{-1}$ where $f_{\it esc}$
is poorly known.  In the supernova wind model
of \citet{taniguchi00}, the required star formation rates are
$\tsim 200 \, \msun \Myr^{-1}$.  This rough agreement between the
required star formation rates of three different models is somewhat
frustrating, with only the model of AGN photoionization allowing
a different rate.  

\section{DISCUSSION}

The most basic, and perhaps the most surprising result of this work,
is that most of the gravitational cooling radiation comes from gas at
$T < 2 \times 10^4 \MK$, far below the typical virial temperatures of
galaxies (see Figure~\ref{fig.coolfrac}).  
In these simulations (and those of Kay et al.\ 2000), most of the
gas that settles into galaxies is never heated to the virial temperature
of a galaxy-mass dark matter halo.  
In \S 5.1 we discuss some
of the numerical issues surrounding this result.
Astronomical implications are discussed in \S 5.2.

\subsection{Numerical Uncertainties}

While numerical simulations are a useful guide to physical intuition,
they are far from a perfect model of reality.  With regard to our
principal result---the temperature distribution of the cooling---it is
easy to think of some limitations of the current code that might
affect cooling estimates.  For example, the omission of molecules and
metals might have some effect.  
The main effect
of molecular cooling would simply be to allow cooling below $10^4\MK$
in the densest regions, so it would not affect our \lya\ predictions.
Metals might channel some of the \lya\ cooling into other lines, and
it would allow more high temperature gas to cool. 
However, the metallicity must be above a few percent of solar before
it significantly affects the cooling curve, and such a high metallicity
seems somewhat unlikely for gas that is falling into galaxies for
the first time at high redshift.  The key feature that leads
to high \lya\ luminosity is that most gas that enters high-redshift
galaxies is never heated to high ($T>10^5\MK$) temperatures at all.
In this regard, a more worrisome concern is inadequate resolution of
shocks, which could allow gas temperatures to rise more slowly than
they should.

Concern about numerical resolution in general can be addressed by
examining the simulations L11/32 and L11/64.  As we mentioned earlier,
L11/32 has the same resolution as our main simulation L144, but it covers
a smaller volume.  L11/64 resimulates the same volume as L11/32 with
the same initial density field, but with eight times the mass
resolution and twice the spatial resolution.  These two resolutions
are compared in Figure~\ref{fig.compare}, where we plot the luminosity
of the cooling radiation emitted in \Lya\ for both simulations. Since
the initial density fields are identical, we can compare the \Lya\ 
cooling luminosity galaxy by galaxy.  We find that although the
individual \Lya\ cooling luminosities can vary, the ratio of \Lya\ 
cooling luminosity in the two simulations scatters around one, giving
us confidence in our ability to predict the \Lya\ cooling luminosity
in the L144 simulation.  The amount of scatter is not surprising in
view of the highly stochastic nature we find for the cooling
luminosity in individual simulated galaxies.

\ifthenelse{\boolean{apj}}{\compareresfig}{}

Though it is encouraging that the cooling luminosity function is
consistent between our two different resolutions, the cooling could be
altered by resolution effects if the important scales are below the
resolution of even the L11/64 simulation.  This caveat may undermine
our prediction that \Lya\ should dominate H$\alpha$ and X-ray
emission, as the presence of stronger shocks would increase collisional
ionization and recombination radiation.
We will eventually be able to test for resolution effects over a wider
dynamic range, but this will require other, more computationally
demanding simulations.

Our method for time integration of the thermal energy also deserves
comment.  As mentioned above, we damp the cooling when the cooling
times are shorter than the minimum timestep.  It would be surprising
if this substantially affects the total radiated energy.  However, if
there is an induced error, it is in the direction of causing us to
slightly overestimate the temperature of the gas and hence slightly
overestimate the importance of X-ray emission relative to \Lya\ 
emission.  

\subsection{Observational Implications}

Laying aside our numerical concerns, which can only be fully addressed
by future work, let us summarize our principal astronomical results.
We find that large amounts of gravitationally induced \Lya\ radiation
should be produced around massive galaxies in the early universe.  The
gravitational cooling radiation is smaller than the stellar UV output, but its
relative significance increases with the galaxy mass, and its physical
extent is quite different, making it less likely to be extinguished by dust. 
The \Lya\ luminosity function extends up
to $\tsim 10^{44} \Merg \Ms^{-1}$, and it reaches a peak at $z \approx 2$.

We suggest that this gravitational cooling radiation could explain the blobs
found by S99.  The luminosities, number densities, and clustering are
consistent with current constraints.  The cooling
radiation is dominated by collisional excitation, so that the expected
ratio of H$\alpha$ to \Lya\ is small.  The size of the \Lya\ emission
region is unlikely to be as large as the observed blobs, but we find
that sizes approaching those observed can be produced by resonant 
scattering of the \Lya\ radiation.

If our model of the blobs is correct, it follows that the currently
observed blobs are merely among the brightest members of their class
at $z=3$.  Our derived luminosity function goes roughly as
$d{\cal N}(>L)/dL \propto L^{-1.5}$ at S99's luminosity threshold.  If
the sky background dominates the pixel
noise, detecting much fainter objects
at $z=3$ will be difficult.  However, our luminosity evolution
suggests the surface brightnesses of the blobs may peak at $z \sim 1$.
Even at $z=2$, we expect $\tsim 6$ times as many blobs at S99's
limiting surface brightness as at $z=3$ (these numbers are
sensitive to cosmology).  Ground-based observations at the lowest achievable
redshift, $z \approx 2$, may be the optimal method for finding these
blobs until the launch of NGST.

The results we find have applications beyond the \Lya\ blobs.  For
example, semi-analytic models of galaxy formation generally assume
that gas in galaxies should be cooling principally at the virial
temperature of the galaxy, and hence be emitted in X-rays.  In the
picture of \citet{white91}, the gas in a halo shock heats to the virial
temperature, then condenses onto the central galaxy from the inside
out as it cools.  The rate of cooling is given by the minimum of the
dynamical gas infall rate and the growth of the mass within the
cooling radius, defined to be the radius where the cooling time equals
the cosmic time.  The emitted cooling is taken to have a near
isothermal spectrum at the virial temperature.

A problem with this picture is that the expected X-ray emission from
individual galaxies has not been detected with ROSAT \citep{benson99}.
One reason for this failure may be that the cooling radiation is not
emitted primarily at the virial temperature but instead covers a large range
of temperatures, with $10^4 \, \MK$
predominant.  This in turn suggests that the assembly of gas to form
galaxies is a more gentle process than in the \citet{white91} picture.  
Once gas
has collected into cool dense objects, it seems that it is difficult
to force it out of that state, and dissipation mostly takes place
through efficient atomic lines.
Dominance of low temperature cooling does not necessarily require
that galaxy assembly proceed by mergers of discrete, cold objects.
If the gas encounters a series of weak shocks as it falls into a
galaxy instead of a single strong shock, it may be able to radiate
its internal energy as quickly as it acquires it, never reaching
high temperature.

During preparation of our paper, a preprint by \citet{haiman00} appeared
proposing a very similar model for the S99 blobs.
We note some of the similarities and
differences of the two models here.  
\citet{haiman00} base their analysis
on a semi-analytic
calculation, using a merger tree formalism and a spherical collapse
model.  In contrast to previous semi-analytic calculations that
assume the energy emerges at the virial temperature, \citet{haiman00}
argue that cooling will be so efficient that most energy will be
radiated at $\tsim 10^4 \MK$.  In general, their picture is in
reasonable agreement with ours, but it differs in quantitative details.  
We find lower amounts of \Lya\
radiation by a factor 2--4, partly because they ignore H~I 2-photon
emission and bremsstrahlung as coolants.  Their treatment of radiative
transfer seems to overemphasize the frequency shift due to resonant
scattering and underestimate the spatial scattering. Because they
assume the \Lya\ radiation is produced out to the virial radius, they
do not consider this a problem.  Our emission is somewhat more
centrally concentrated.  They assume monolithic collapse, so that for
a given galaxy there is first a cooling stage and then a star
formation stage.  In contrast, we find that cooling and star
formation happen simultaneously in massive galaxies.

Finally, \citet{haiman00} find that the efficiency of star formation in 
producing
\Lya\ radiation is only twice that due to gravitational cooling.  In
our simulation this ratio is about 20.  The difference comes in part
from their channeling all the cooling radiation through \Lya\, and in
part from their assumption about the importance of feedback.  They
assume that only 10\% of the gas forms stars.  Since at the present day
most galaxies are dominated by stars rather than gas, this implies the
remaining gas must be blown out again by galactic winds.  In our
simulations, the efficiency of forming stars out of eligible gas is
only 10\% at each timestep, but the gas is not blown out, and eventually
most of the cooled gas is turned into stars.  The effectiveness of
feedback in removing gas from galaxies is currently a major question
in astrophysics, and the comparison of cooling radiation and stellar
emission may eventually help to constrain the answer.

Further theoretical work is needed to test the robustness of 
our predictions over a wider dynamic range of numerical parameters.
Further observational work is needed to test whether the 
predicted \lya\ emission from young galaxies exists in the real universe.
This emission, perhaps already observed in the form of the \Lya\
blobs, represents a novel form of radiation from galaxies that
potentially offers a direct view of the process of galaxy formation.

\acknowledgments

We thank Chuck Steidel, Martin Weinberg, Chigurupati Murali, 
Paul Francis, and Mark Voit
for helpful discussions.
This work was supported
by NASA Astrophysical Theory Grants NAG5-3922, NAG5-3820, and NAG5-3111,
by NASA Long-Term Space Astrophysics Grant NAG5-3525, and by the NSF under
grants ASC93-18185, ACI96-19019, and AST-9802568.
The simulations were performed at the San Diego Supercomputer Center,
NCSA, and the NASA/Ames Research Center.

\ifthenelse{\boolean{apj}}{}{\clearpage}

\ifthenelse{\boolean{apj}}{}{\coolfracfig}

\ifthenelse{\boolean{apj}}{}{\lfcumfig}

\ifthenelse{\boolean{apj}}{}{\lyalfzevlnfig}

\ifthenelse{\boolean{apj}}{}{\lyaimagefig}

\ifthenelse{\boolean{apj}}{}{\extentfig}

\ifthenelse{\boolean{apj}}{}{\bmassfig}

\ifthenelse{\boolean{apj}}{}{\sfrfig}

\ifthenelse{\boolean{apj}}{}{\epotfig}

\ifthenelse{\boolean{apj}}{}{\compareresfig}

\end{document}